\documentclass[lettersize,journal]{IEEEtran}
\usepackage{amsmath,amsfonts}
\usepackage{algorithmic}
\usepackage{array}
\usepackage[hidelinks]{hyperref}
\usepackage[caption=false,font=normalsize,labelfont=sf,textfont=sf]{subfig}
\usepackage{textcomp}
\usepackage{stfloats}
\usepackage{url}
\usepackage{verbatim}
\usepackage{graphicx}
\usepackage{adjustbox}
\def\BibTeX{{\rm B\kern-.05em{\sc i\kern-.025em b}\kern-.08em
    T\kern-.1667em\lower.7ex\hbox{E}\kern-.125em}}
\usepackage{balance}
\begin{document}
\bstctlcite{IEEEexample:BSTcontrol}	
	
\title{Towards Domain Generalization for ECG and EEG Classification: Algorithms and Benchmarks}
\author{Aristotelis Ballas and Christos
	Diou, \IEEEmembership{Member, IEEE}
\thanks{Manuscript received 18 November 2022; revised 2 May 2023; accepted 22 June 2023. The work leading to these results has received funding from the European Union’s Horizon 2020 research and innovation programme under Grant Agreement No. 965231, project REBECCA (REsearch on BrEast Cancer induced chronic conditions supported by Causal Analysis of multi-source data).}}

\markboth{PREPRINT. Accepted in: IEEE Transactions on Emerging Topics in Computational Intelligence, June~2023}%
{How to Use the IEEEtran \LaTeX \ Templates}

\maketitle

\begin{abstract}
Despite their immense success in numerous fields,
machine and deep learning systems have not yet
been able to firmly establish themselves in mission-critical applications in
healthcare.
One of the main reasons lies in the fact that when models are presented with
previously unseen, Out-of-Distribution samples, their performance 
deteriorates significantly. This is known as the Domain
Generalization (DG) problem. Our objective in this work is to propose a
benchmark for evaluating DG algorithms, in addition to introducing a novel
architecture for tackling DG in biosignal classification. 
In this paper, we describe the Domain Generalization problem for biosignals, focusing on electrocardiograms (ECG) and electroencephalograms (EEG) and propose and implement an open-source biosignal DG evaluation benchmark. Furthermore, we adapt state-of-the-art DG algorithms from computer vision to the problem of 1D biosignal classification and evaluate their effectiveness. Finally, we also introduce a novel neural network architecture that leverages multi-layer representations for improved model generalizability. By implementing the above DG setup we are able to experimentally demonstrate the presence of the DG problem in ECG and EEG datasets. In addition, our proposed model demonstrates improved effectiveness compared to the baseline algorithms, exceeding the state-of-the-art in both datasets. Recognizing the significance of the distribution shift present in biosignal datasets, the presented benchmark aims at urging further research into the field of biomedical DG by simplifying the evaluation process of proposed algorithms. To our knowledge, this is the first attempt at developing an open-source framework for evaluating ECG and EEG DG algorithms.
\end{abstract}

\begin{IEEEkeywords}
	Biosignal classification, deep learning, domain generalization, 1D signal classification,
	electrocardiogram (ECG) classification, electroencephalogram (EEG) classification
\end{IEEEkeywords}

\section{Introduction}
\label{sec:introduction}
%\IEEEPARstart{W}{hat} are human beings if not adaptable? "The evolution of the brain is the most obvious
%example of how we evolve to adapt," said Rick Potts, director of the Human Origins Program at the
%Smithsonian Institution National Museum of Natural History \cite{masseyclimatewire_humans}. The plasticity
%of the human brain has allowed us to adjust to all kinds of environments and interactions, throughout history.
%Even from a very young age, humans are able to digest their world into robust representations, from which
%they learn and figure out ways to navigate in new environmets. For example, after seeing an aeroplane in
%real-life for the first time, or even in photographs, a child will most likely be able to recognize it in most of the cases,
%regardless if it's in the air, on ground, in front of a mountain, etc. Unfortunately, machine learning (ML) algortihms
%have yet to produce models with similar traits. The holy grail of ML is to provide models which can
%learn general and predictive knowledge from initially unstructed data. However in most cases, since most ML models
%are trained based on the i.i.d assumption, meaning that both training and testing data are identically and
%independently distributed, they inadvertently tend to incorporate statistical dependecies present in said data.

%%%%%%%%%%%%%%%%%% Alternative Intro %%%%%%%%%%%%%%%%%%%
\IEEEPARstart{D}{omain} generalization is a fundamental problem in machine
learning (ML) today \cite{wang2022generalizing}. Despite the fact that deep
learning (DL) \cite{lecun2015deep} models have seen immense success in the past
few years \cite{he2016deep, krizhevsky2017imagenet, vaswani2017attention}, even
surpassing experts in some cases \cite{he2015delving, mnih2015human,
	mckinney_international_2020}, they often fail to mimic the adaptability
prowess of humans.  The development of highly generalizable and robust ML
models proves to be exceptionally difficult in numerous cases, as the
distribution shifts present across separate datasets or databases cause a
model's performance to deteriorate \cite{recht2019imagenet}, or even completely
break down \cite{zhou2022domain}, when evaluated on previously unseen data.

\begin{figure}[h]
	\centering
	\includegraphics[width=0.8\columnwidth]{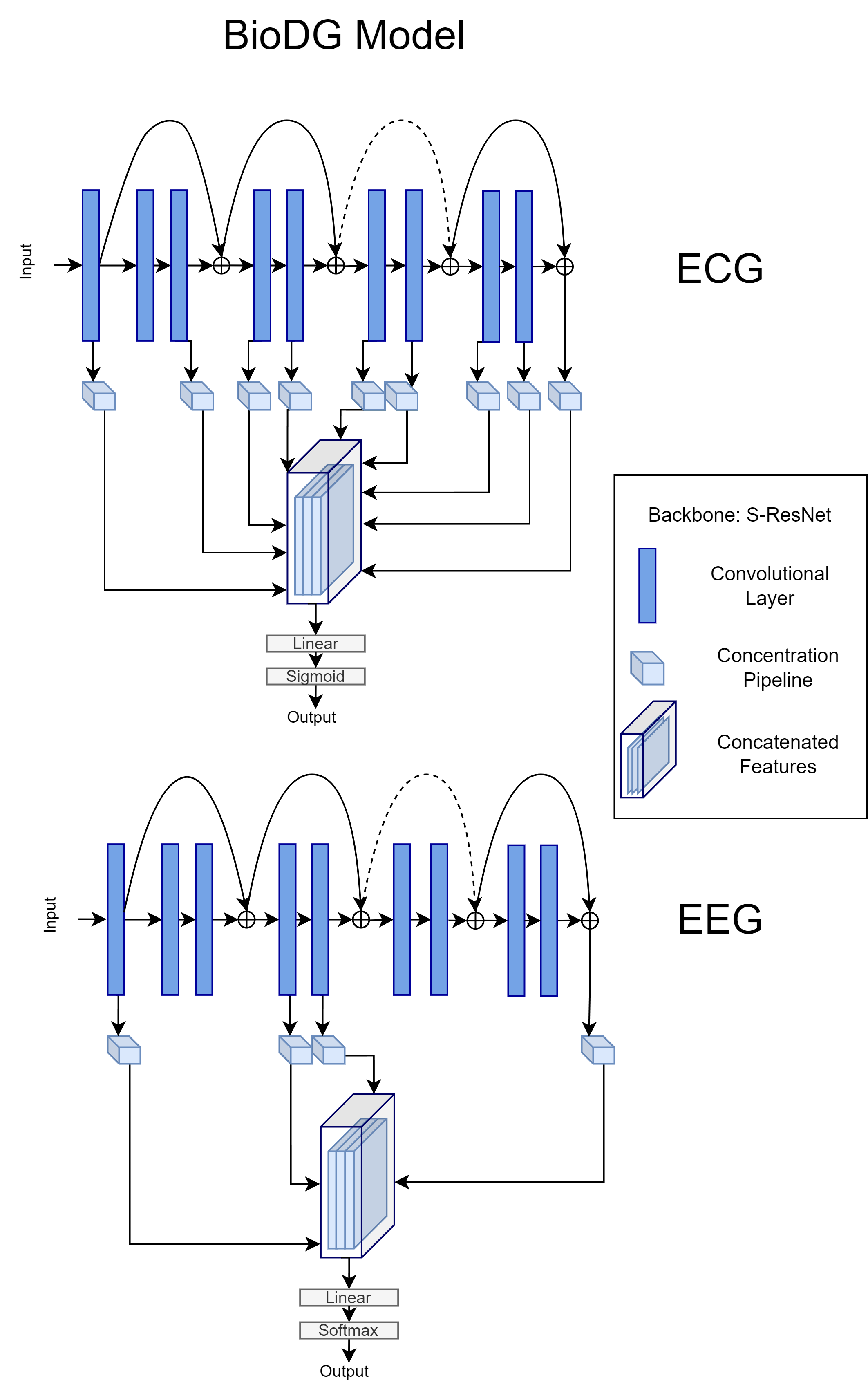}
	\caption{Visualization of the alternative architectures for the ECG and EEG classification experiments. The backbone of both models is a small ResNet or S-ResNet (Fig. \ref{fig:backbones}) with 13 convolutional layers (including the conv layers in the residual connections) in total. In the ECG model, we extract features from a total of 8 intermediate layers and concatenate them to the backbone network's output, as shown in the above figure. In the EEG model, we connect concentration pipelines to 3 intermediate layers and concatenate the processed features to the backbone network output.  
	The solid lines represent normal residual connections between CNN layers where the passed features retain their dimension, while the features passed through the dashed lined connections are downsampled to match the dimension of the previous layer output.}
	\label{fig:model}
\end{figure}

Most ML models depend on the assumption that the test samples are
independent from and identically distributed (i.i.d.) with the training data. In
practice, the independence assumption commonly holds, however the test data
often follows a different data generating distribution than the training
data. These differences are often the result of contextual changes e.g., in the
background of images (for natural images) or the use of different medical
imaging equipment (for medical images). ML models tend to incorporate
statistical correlations present in their training data, even if these are
spurious, in the sense that they do not convey useful information about the
problem at hand. The fact that these correlations may not be present (or may
manifest differently) in the test data, leads to significant degradation in
performance when DL models are applied in practice.

This is one of the main reasons why DL models have not been able to establish
themselves in production and several mission-critical applications, including
healthcare settings. This problem was firstly introduced in the ML
literature as \emph{domain generalization} (DG)
\cite{blanchard2011generalizing}. DG emphasizes on developing algorithms which
are not affected by the distributional shift present in distinct data domains
of the same problem, by not incorporating spurious correlations in their
representations. DG models should ideally maintain their performance among
different training or \emph{source} and test or \emph{target} data domains of
the same problem.

Lately, an abundance of works have been proposed towards tackling the DG
problem. Nonetheless, although noteworthy progress has been made mostly in
computer vision, the field currently lacks in researching methods for DG in
one-dimensional (1D) signals. To this end, in this work we turn our attention
to biosignals and specifically to 1D signals originating from electrodes:
electrocardiograms (ECG) and electroencephalograms (EEG). Due to their medical
nature, biosignal classification renders the need for
highly robust and generalizable algorithms of the utmost importance. In a field
of constantly evolving equipment and screening methods, AI-enhanced clinical
support systems must gain the trust of medical practitioners by maintaining
their performance against diverse data distributions. Additionally, most
biosignal datasets suffer from class imbalance, creating a need for models with
the ability to identify rarely represented categories (i.e., diseases). Our aim
in this work is to encourage future research in biosignal domain
generalization, by describing and experimentally demonstrating the DG problem
in biosignals and by developing an open-source evaluation benchmark based on
publicly available biosignal datasets. Specifically, we look into the
classification of 12-lead ECGs and 62-channel EEGs. The main contributions of this paper are the following:
\begin{itemize}
	\item We introduce a DG evaluation benchmark, namely \emph{BioDG}\footnote{
		Code is available at \url{https://github.com/aristotelisballas/biodg}.}, for 12-lead 
	ECG and 62-channel EEG biosignals.
	For our experiments we use the 6 datasets from PhysioNet's \cite{alday_classification_2020} public database, for the ECG signals,
	and the SEED \cite{duan_differential_2013, zheng_investigating_2015},
	SEED-FRA \cite{schaefer_assessing_2010, liu_identifying_2022} and SEED-GER
	\cite{schaefer_assessing_2010, liu_identifying_2022} for the EEGs. To our
	knowledge, this is the first attempt at developing an open-source and
	reproducible evaluation framework, combining both ECG and EEG
	signals. Moreover, in this setup we follow the classic
	\emph{leave-one-domain-out} DG protocol \cite{Li_2017_ICCV} and move past
	the leave-one or leave-multiple subjects-out protocols described in previous works.
	\item We experimentally exhibit the distributional shift present in the above
	datasets, thus validating our claim that additional DG research should be
	poured into the medical AI field.
	\item We adapt and implement state-of-the-art DG computer vision algorithms
	for 1D data and experiment on our proposed DG evaluation setting.
	\item We build on our previous work \cite{ballassetn2022, 9898255} and
	propose a DG neural network architecture which takes advantage
	of representations from multiple layers of a Convolutional Neural Network and exhibit its capabilities against state-of-the-art DG methods, in both ECG and EEG classification.
\end{itemize}

\section{Background and Terminology}
Let $\mathcal{X}$ and $\mathcal{Y}$ be a nonempty input and output space
respectively. A \emph{domain} $D$ is a composition of sample and label pairs $(x, y)$ in $(\mathcal{X}, \mathcal{Y})$, drawn from the (unknown) data distribution $P_{XY}$. In contrast to fully supervised learning,
in which the common assumption is that both training and evaluation data are identically distributed, Domain Generalization algorithms aim to learn a parametric model $f(\cdot; \theta)$ trained on samples $(\mathbf{x}^{(s)}, y^{(s)})$ drawn from $N$  \emph{source} ($N>1$) domains, which is able to generalize to $K$ ($K>1$) unseen \emph{target} domains.

In the context of DG, classification problems can be grouped into two main categories \cite{zhou2022domain}, namely \textit{multi-source} and \textit{single-source}. Multi-source DG aims at training models which are aware of the distinction between multiple but related source domains $(N>1)$. As a result, the goal is to learn representations which remain unaffected by the distribution shift between the marginal distributions of each source domain. On the other hand, the presence of several underlying data domains is irrelevant to single-source DG algorithms, as they are trained under the assumption that all training data is sampled from a single distribution $(N=1)$. Single-source DG algorithms can therefore be labeled as domain-agnostic.

In the current work, we attempt to improve the generalization ability of a
model $M_\theta$: $\mathcal{X}$ $\rightarrow$ $\mathcal{Y}$ to detect
domain-invariant attributes of 1D ECG and EEG signals and minimize the affect
of the distributional shift present in distinct biosignal datasets. Since our method is not aware of the presence of separate data domain distributions, it falls under the category of \textit{single-source} DG algorithms.

\section{Related Work}\label{sec:relatedwork}

\subsection{Methods for Improving Generalizability in Biosignal
	Classification}\label{sec:nonDG}
Although only a limited number of works have dealt directly with the problem of
domain generalization for biosignal classification, there have been several
attempts to implicitly address the problem. In this section, we discuss
the state-of-the-art and most noteworthy non-DG methods proposed for the
improvement of the generalization ability of ML algorithms and mitigation of
domain-shift.

\textbf{Domain Adaptation} (DA) \cite{pmlr-v37-ganin15} methods are the most
common approach for enabling a model to generalize to new domains and are perhaps the closest to DG. DA algorithms leverage pre-trained models and use their feature extraction capabilities, by fine-turning them on unseen data distributions or target domains in order to improve model generalizability. Although similar, the difference between DA and DG is apparent as DA models update their parameters based on data drawn from target domains and are evaluated on the same target distributions, whereas DG algorithms hold no prior information regarding the target domains. In \cite{WANG2021339} the authors propose using a cluster-aligning loss, coupled with a cluster-maintaining loss, in order to respectively align the training and test data distributions and reinforce the discriminative ability of their model, for the classification of arrhythmia heartbeats. The authors of \cite{weimann2021transfer} demonstrate the effectiveness of pre-training vanilla convolutional neural networks (CNN) on large public raw ECG datasets, which are then fine-tuned for the classification of heart arrhythmias as well, finding a significant improvement in the downstream classification task. A plethora of DA methods have also been proposed in EEG processing. In an attempt to produce a model which generalizes across two different emotion recognition datasets, \cite{8337789} investigates the implementation of several DA algorithms, such as MIDA \cite{yan2017learning}, TCA \cite{pan2010domain}, SA \cite{fernando2013unsupervised} and others. The authors report an overall improvement over the baseline model. In turn, the authors of \cite{9154600} employ an end-to-end deep DA method which leverages a domain adversarial discriminator to match the distribution shift between source and target data, in a motor imagery classification problem.

\textbf{Self-Supervised Learning} (SSL) \cite{misra2020self} methods have also
proven effective in boosting a model's generalizability, as the main goal is
for the model to learn domain-invariant representations by discriminating
transformed or augmented input data from the initial signal.
A first example in ECG classification is \cite{MEHARI2022105114}, in which the
use of contrastive predictive coding produces a model with increased accuracy
against its fully supervised counterpart, along with some level of robustness
against physiological noise. Moreover, in \cite{ballascinc2022} the authors
investigate the effectiveness of SSL for heart murmur detection,
while also explore the most effective input data augmentations. SSL 
has also been implemented for emotion recognition from ECGs in \cite{sarkar2020self}, 
where the authors report state-of-the-art results for the 
respective dataset. As far as EEG is concerned, several works tackling sleep 
stage classification and pathology detection \cite{8918693, banville2021uncovering, 9385345}, 
attempt to extract representations and the underlying structure of the 
physiological signals signals via contrastive learning.

\textbf{Attention}-based architectures \cite{vaswani2017attention} seem to be
effective and successful in improving the generalization ability of biosignal
classification models as well. The use of handcrafted ECG features paired with
a transformer network is explored in \cite{9344053}, for the classification of
cardiac abnormalities, yielding promising results. Furthermore, in
\cite{9391844} the authors apply several attention-based models and report
improved results on a public motor movement/imagery dataset, while
\cite{9684393} uses a similar model to hierarchically learn the discriminative
spatial information from electrode level to brain-region level.

\begin{figure}[tbp]
	\centering
	\includegraphics[width=\linewidth, ]{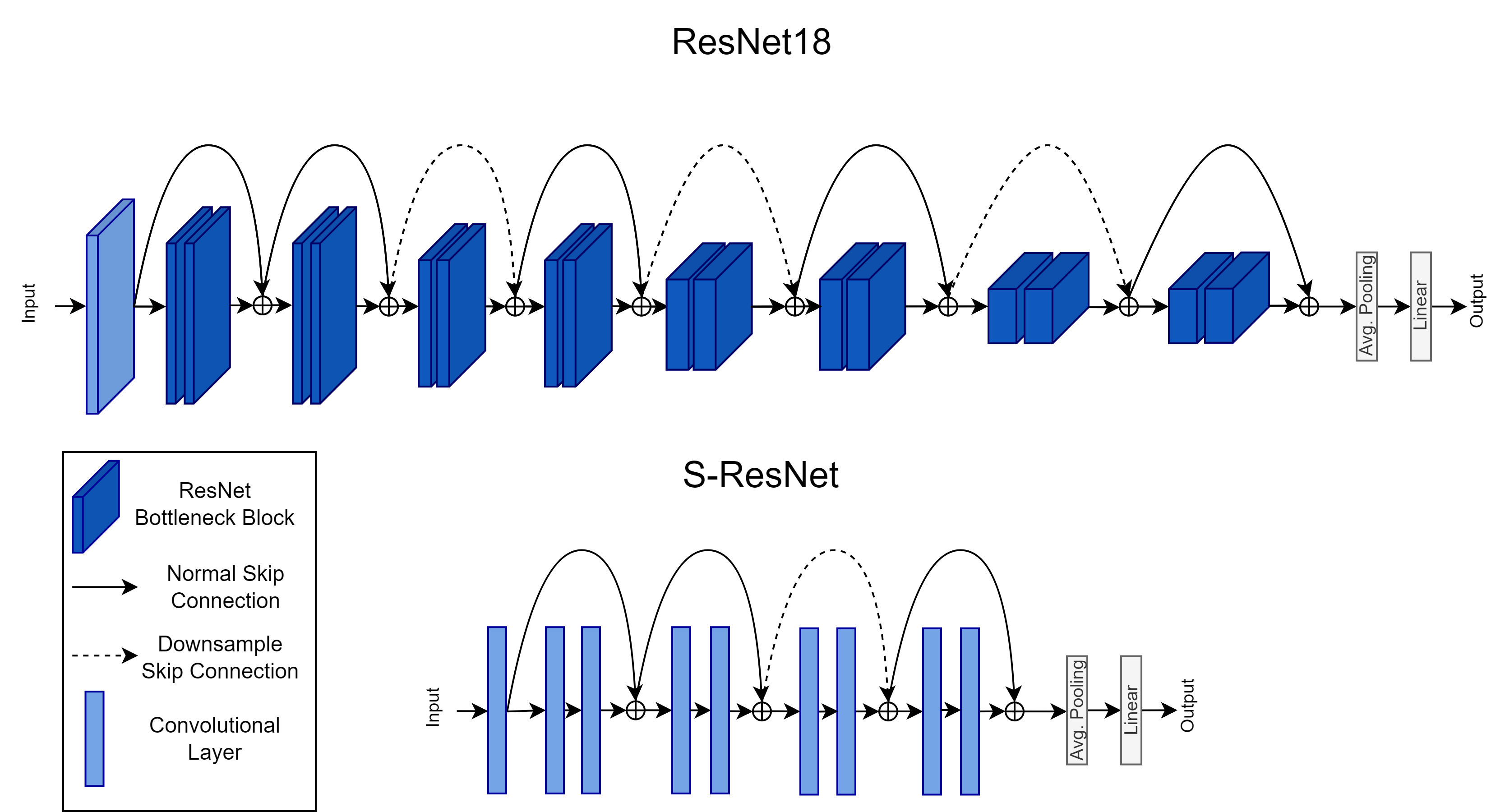}
	\caption{Visualization of the backbone networks used in our experiments. The
		top network is a standard ResNet-18. The bottom network is a smaller
		ResNet or S-ResNet, consisting of 4 residual blocks, with 2 convolutional layers each. Therefore, the S-ResNet consists of a total of 9 convolutional layers. The lines above the network blocks and conv layers represent ResNet skip connections. The solid lines indicate that the passed feature maps retain their dimension, while the feature maps passed through the dashed connection lines are downsampled to match the dimension of the previous layer output. Both networks are adapted for 1D signal classification.}
	\label{fig:backbones}
\end{figure}

\subsection{Domain Generalization Methods in Biosignal Classification}\label{sec:dgBiosignals}
Even though the prior work in biosignal DG is limited, researchers have shown
some interest in the past. For example, the authors of \cite{9298838} propose a
DG setup for normal vs abnormal phonocardiogram classification and claim their
ensemble classifier fusion method yields significant accuracy improvement
across domains. In \cite{9344210} multi-source domain-adversarial training
\cite{ganin2016domain} is implemented to overcome the heterogeneity present in
ECG signals, but no results are presented between source and target domains.
The same follows in \cite{ma2019reducing}, where the authors propose an
adversarial domain generalization framework to reduce variability between
signals originating from distinct subjects but not from different datasets (domains). With the exception of \cite{9298838}, all previously proposed DG biosignal papers formulate each patient as a separate domain and aim to improve cross-subject generalization in their models. However, in each respective dataset the data distribution remains similar (e.g. population groups with matching characteristics, same hospital with specific equipment) and the produced models are evaluated on comparable data originating from similar data generating processes. DG algorithms take into consideration the above and focus on producing robust models which are able to extract domain-invariant representations from input signals and generalize to completely unseen data distributions \cite{wang2022generalizing}.
In the sections to come, after describing methods similar to our own, we discuss DG algorithms which derive from computer
vision, adapt them accordingly and explore their effectiveness on 1D biosignal data.

\subsection{Multi-Layer Representation Learning}
Several aspects of multi-layer representation learning have been researched in the machine learning literature, from which we drew inspiration for our proposed method. In their paper \cite{hariharan_hypercolumns_2015}, Hariharan et al. propose to utilize features throughout multiple layers in a CNN, to build Hypercolumns for image segmentation. However, to build Hypercolumns an initial set of bounding boxes indicating the point of interest on the image is required. Hypercolumns have been also adopted in the medical image domain, where the authors of \cite{togacar_enhancing_2021} use them to detect stages of Alzheimer's disease. In addition, Feature Pyramid Networks (FPNs) stack representations from several layers of the network in a pyramid-like manner and were proposed for object detection in images. The pyramids are constructed by pooling 7$\times$7 features from provided regions of interest and are passed through two hidden fully connected layers before the final classification. Another approach in multi-layer representation learning is feature reuse. For example, the authors of \cite{huang_densely_2017} propose connecting all layers with matching feature-map sizes via dense connections to tackle the vanishing gradient problem and report improved performance in several image object recognition benchmarks. Recently, \cite{ballasdiouattention2023} proposed employing self-attention mechanisms to image feature maps from multiple layers of a CNN, to push the model to attend to class-relevant cross-channel attributes. Finally, U-Nets \cite{ronneberger_u-net_2015} were proposed for improved medical image segmentation, in which extracted feature maps from early layers of a CNN are upsampled and concatenated to deeper layers of the network.

In contrast to previous methods designed for image processing, our multi-layer representation algorithm is proposed for 1D biosignal classification. The novelty of our method lies in the fact that each level of information extracted is processed independently, without getting inserted back into the network and potentially polluted with representations corresponding to spurious correlations. As a result, we hypothesize that a classifier will be able to base its predictions on the representations which remain invariant between different domains.

\section{Adapting Computer Vision Domain Generalization Methods to Biosignal Classification}\label{sec:dgGeneral}

Domain Generalization methods focus on maintaining high accuracy across both known and unknown data distributions. The core 
difference between DG and other paradigms is the fact that during training, the model has no access to the
test distributions whatsoever, which makes it a significantly more difficult problem than DA. 
In the ML literature, most DG methods have been proposed for computer vision tasks. Therefore, for our evaluation we select
to use as baselines the most widely accepted algorithms, after adapting them for 1D biosignal classification. 

\begin{figure*}[htbp]
	\centering
	\includegraphics[width=\linewidth, ]{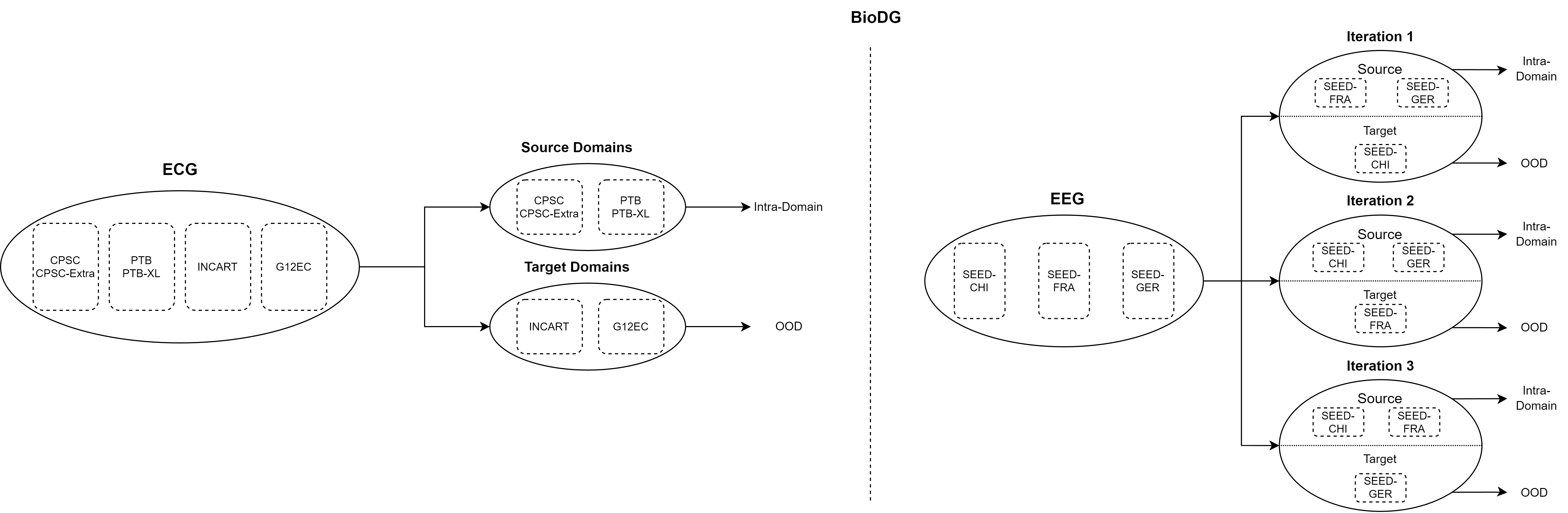}
	\caption{Illustration of the proposed BioDG benchmark, for ECG and EEG classification. 
		Each dataset is split into Source and Target domains. The models are trained only on source domain data and are then evaluated on both held-out intra-distribution (drawn from Source domains) and OOD (drawn from Target domains) data.}
	\label{fig:benchmark}
\end{figure*}

To push a model to learn invariances related to causal structures of a class and therefore enable out-of-distribution generalization, 
\cite{arjovsky_invariant_2020} introduces invariant risk minimization or IRM. IRM is used in order to estimate invariant correlations
across numerous training distributions and to learn a data representation for which an optimal classifier
matches throughout all training distributions. In the context of 1D biosignal classification, IRM could push a model to learn the parts or
attributes of the signal, which remain invariant between different datasets.
The authors of \cite{Zhang_2021_CVPR} 
use Random Fourier Features and sample weighting to compensate for the complex, non-linear correlations amongst non-iid 
data distributions, while JiGen \cite{Carlucci_2019_CVPR} proposes solving jigsaw puzzles via a self-supervision task, in an attempt to restrict semantic feature learning. Specifically, JiGen's objective is split into two parts. First, the model attempts to minimize the error between predicted and true label and secondly, tries to reconstruct the permuted, decomposed and split-into-patches source images. Additionally, SagNets \cite{nam2021reducing} aim to reduce domain gap in images by disentangling their style encodings.

Meta-learning approaches have also been proposed \cite{li2018learning, pmlr-v70-finn17a}. In a recent work, \cite{10.1007/978-3-030-58607-2_12} leverages the variational bounds of mutual information in the meta-learning setting and uses episodic training to extract invariant representations. 

In \cite{huangRSC2020}, the authors introduce a self-challenging training heuristic (RSC) that discards representations associated with the higher gradients of a neural network. They hypothesize that the dominant features present in the dataset are associated with their model's activations with higher gradients. Under this scope, they demonstrate that by discarding the higher activations, and therefore by iteratively challenging the model, the network is forced to activate the remaining features which should correlate with the data labels. A similar trend should hold in signal classification, as the network should be able to disregard features attributing to noise or distributional shift.

The authors of \cite{10.1007/978-3-030-58542-6_5} combine batch
and instance normalization to extract domain-agnostic feature representations, in contrast to \cite{venkatesh2020calibrate}, which explores the use of 
data augmentation techniques in the DG setting. 
Furthermore, \cite{li2018domain} extends adversarial autoencoders with a maximum mean discrepancy (MMD) measure for the alignment of different domain distributions and also uses adversarial feature learning
to match the aligned distributions to an arbitrary prior distribution. 
Finally, correlation alignment or CORAL \cite{sun2016return} 
was extended in \cite{sun2016deep} for aligning correlations of DNN layer activations and addressing performance degradation 
due to domain shift.

\begin{table}[h]\centering 
	\begin{center}
		\caption{Training parameters for the baseline methods evaluation. On the left
			column, we summarize the parameters for each of the backbone networks
			implemented for the ECG dataset and on the right for the EEG dataset.}
		\label{tab:baseline_params}
		\begin{adjustbox}{width=1\linewidth}
			\begin{tabular}{|c||cc||c|}
				\hline
				%\multicolumn{4}{|c|}{\hspace{0.4cm}Dataset \hspace{1.9cm} ECG \hspace{1.7cm} EEG} \\
				Dataset & \multicolumn{2}{c||}{Intra-Distribution} & {OOD} \\
				\hline
				% Dataset & ~ & ECG & EEG \\
				%\hline
				Backbone & ResNet-18 & S-ResNet & S-ResNet \\
				\hline
				Optimizer & Adam & Adam & Adam \\
				Learning Rate & 0.001 & 0.001 & 0.00009 \\
				Weight Decay & 0.0005 & 0.0005 & 0.00005 \\
				Batch Size & 128 & 128 & 64\\
				LR Decay Epoch & 24 & 24 &14\\
				%\\[-1.ex]
				Epochs & 30 & 30 & 20\\
				\hline
			\end{tabular}
		\end{adjustbox}
	\end{center}
\end{table}

Unfortunately, not all of the above methods can be adapted to 1D biosignal
classification. The abundance of the algorithms either need to have knowledge about the number of separate domains (multi-source DG) in the training data or are based on image style transfer. Therefore, we elect to adapt a subset of the top performing \textit{single-source} DG algorithms for our experiments. As they were initially proposed for CV, their backbone networks consisted of vanilla ResNet-18 or ResNet-50 residual networks
\cite{he2016deep}. As a first step, we converted all 2D layers to 1D layers in
order to be able to train them on the available ECG and EEG signals and
experimented with a ResNet-18 backbone network. For the ECG classification task
we found that the ResNet-18 was able to achieve good convergence. However, due to the limited number of signals, and therefore a small overall dataset size (more in Sections \ref{sec:datasets} and \ref{sec:experimental_setup}), it was not able to succeed in converging on the EEG data and thus experimented
with a smaller network with fewer parameters. We therefore propose to use a
smaller custom ResNet, or S-ResNet consisting of 4 residual blocks, with 2
convolutional layers each. Both backbone networks are depicted in Figure
\ref{fig:backbones}. To summarize, for our implementation we adopt the baseline single-source DG methods of CORAL, IRM, MMD and RSC, along with a classic fully supervised model based solely on Empirical Risk Minimization (ERM) \cite{vapnik1991principles, gulrajani2021in}. In addition, the ``S-ResNet'' and ``ResNet-18'' notations, reflect the backbone architecture implemented in each experiment. For the ECG dataset we use a 1D ResNet-18 and a 1D S-ResNet with randomized weights for backbone networks. We train all models for 30 epochs and use a batch size of 128 signals.  We also adopt the Adam optimizer, with a weight decay of $0.0005$ and set the initial learning rate to 0.001. After 24
epochs, the learning rate decays by 0.1 for each remaining epoch. For the EEG
classification task, we implement a S-ResNet with randomly initialized weights
and adapt all available algorithms to use it as a backbone. For the
above setting we train all models for 20 epochs, again using the Adam
optimizer, with an initial learning rate of 9$e-$05, decaying at epoch 14 with
a 0.1 rate and a batch size of 64.  All the above training parameters are
summarized in Table \ref{tab:baseline_params} and all additional hyperparameters
for each DG algorithm were set as described in their respective papers. All methods were implemented using PyTorch \cite{NEURIPS2019_9015} and all models were trained using an NVIDIA RTX A5000 GPU.

\begin{figure}[t]
	\centering
	\includegraphics[width=.8\linewidth, ]{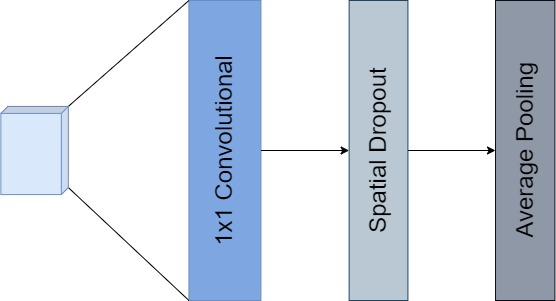}
	\caption{Visualization of the proposed concentration pipeline.
		To transform the extracted feature maps into compressed representations of the initial signal, each concentration pipeline consists of 3 sequentially connected layers: A 1x1 Conv layer, a Spatial Dropout layer and an Average Pooling layer. We hypothesize that these representations
		could contain disentangled attributes of the input data, invariant to the distributional shift present between domains.}
	\label{fig:pipeline}
\end{figure}

\subsection{A Concentration Pipeline for DG in Biosignal Classification}\label{sec:ourMethod}
In addition to the models adapted from the computer vision DG literature, we
also propose an alternative approach towards tackling DG in biosignals. To be
more concise, after producing promising results in computer vision
\cite{ballassetn2022}, we adapt our method for biosignals by building upon our
preliminary work in \cite{9898255}. In this section, we present our proposed
method.

Our approach lies somewhat between the aforementioned DG methods. In most
cases, the above algorithms are implemented with networks which contain
convolutional layers. 
%We hypothesize that it is difficult to avoid
%\emph{entangled} representations, i.e. representations containing both
%class-invariant information and domain-specific features when inferring based
%on representations from the last layers of a deep CNN. 
We hypothesize that a model is unable to infer based on \textit{entangled} representations, i.e representations containing both class-invariant information and domain-specific features, when relying only on the final layers of a deep CNN. Consequentially, the model is prone to base its predictions on spurious correlations present in its training data. We argue that the distribution shift problem between data drawn from unknown domains can be mitigated by leveraging information passed throughout the network. Hence, we propose building representations from features extracted from several layers of a CNN.

\begin{figure*}[htbp]
	\centering
	\includegraphics[width=\linewidth, ]{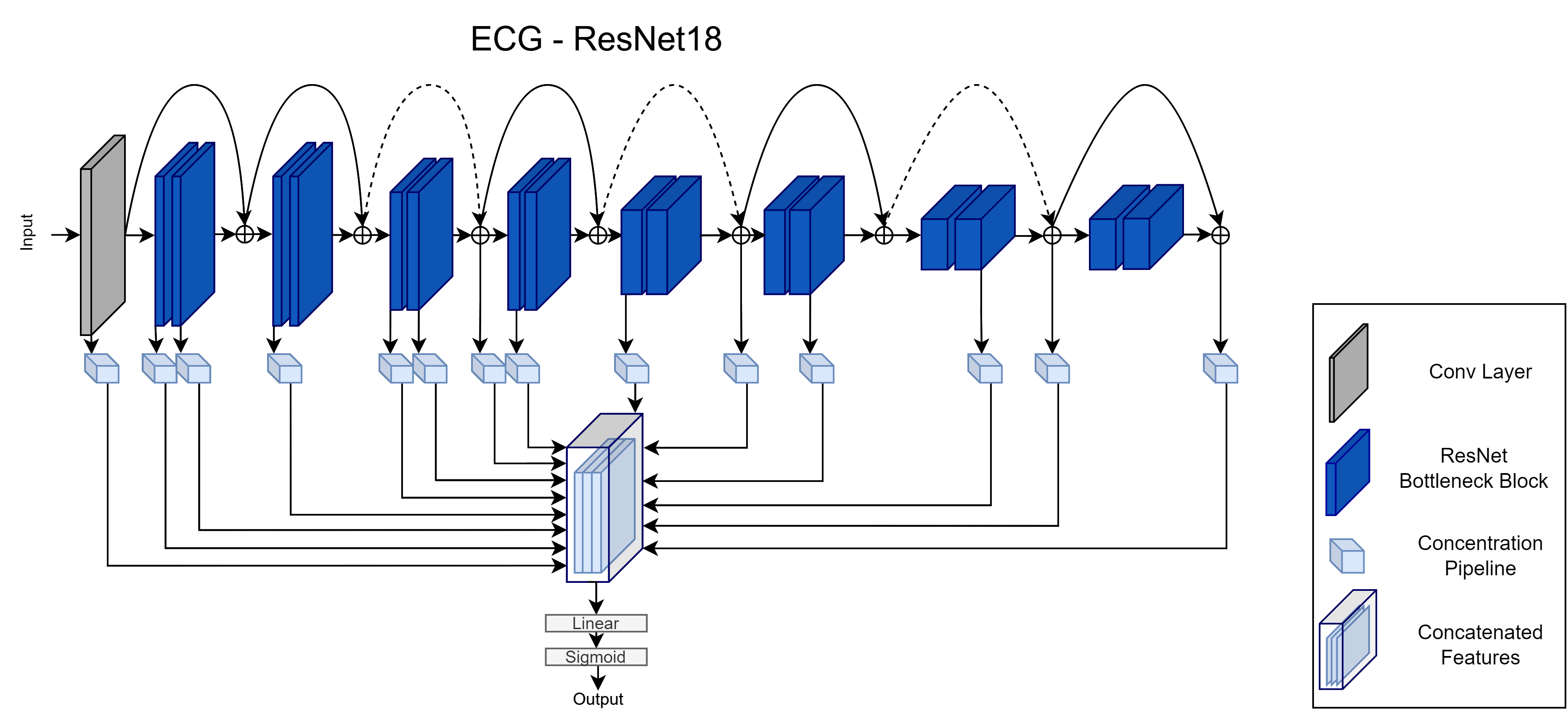}
	\caption{Visualization of our proposed DG neural network architecture. Our method utilizes a standard ResNet-18 as the backbone model. We hypothesize that the extraction and combination of feature maps from multiple layers of a CNN, enables the model to disentangle the input data and infer based on the invariant attributes of the signal. In this implementation we select to connect extraction blocks to 14 different location throughout the backbone network, as shown in the figure. In addition to extracting features from convolutional layers, we also choose to process features extracted from skip connections. As in Figure \ref{fig:backbones}, the solid lines represent normal connections between network layers where the passed features retain their dimension, while the features passed through the dashed lined connections are downsampled to match the dimension of the previous layer output.}
	\label{fig:resnet18}
\end{figure*}

The extraction of these features is accomplished by attaching a
custom sequential \emph{concentration pipeline} (Figure \ref{fig:pipeline}) of layers to multiple levels of the corresponding backbone model. To demonstrate the proposed architecture, we use the same backbones as the above DG methods
and choose to extract features from a total of 15, 9 and 5 layers across the
ECG ResNet-18, ECG S-ResNet and EEG S-ResNet respectively, as depicted in
Figures \ref{fig:model} and \ref{fig:resnet18}. After experimenting with the position and number of connected concentration pipelines we found that for the ECG task, the performance of our model improved when features where extracted from across the network and the majority of the intermediate layers. In contrast, for the EEG experiments we found that less extracted features were needed to boost the model's generalization ability.  Furthermore, we implement the same training parameters as in Table \ref{tab:baseline_params}, for the respective backbone and dataset.

The proposed pipeline contains a 1x1 Convolutional layer, a Spatial Dropout layer and an Average Pooling layer. Each component of the pipeline is implemented for reducing the dimensionality of the extracted feature maps and hopefully yielding invariant attributes of the unstructured input data. Initially, the extracted features are projected into a lower dimensionality by the 1x1 conv layer, leading to a compressed representation of the signal. The spatial dropout layer promotes the independence of the compressed feature maps and acts as a regularizer, while the average pooling layer extracts the average level of information present in the compressed features. Finally, the output vectors of each pipeline are concatenated and passed through the network's classification head, which consists of a fully connected linear layer and the appropriate activation layer. Moreover, as our method has no knowledge of any domain labels whatsoever, it can be categorized as a \textit{single-source} DG algorithm. For simplicity, the above implementations will hereafter be referred to as ``BioDG baseline''.

\section{Evaluation Benchmark}\label{sec:benchmark}
Our primary contribution in this work, is an open-source evaluation benchmark
for DG algorithms in ECG and EEG classification.  We argue that a generalizable
model should be able to maintain its performance across data originating from
several experimental settings, equipment and populations with different
characteristics. For our experiments we select 6 ECG datasets from PhysioNet's public database \cite{alday_classification_2020} and 3 of the SEED
\cite{duan_differential_2013, zheng_investigating_2015, schaefer_assessing_2010, liu_identifying_2022} EEG datasets. In the following
subsections we will describe the experimental setup for each biosignal dataset,
from data preprocessing, to data separation, model training and ultimately
model evaluation.

\subsection{Datasets}
\label{sec:datasets}
The PhysioNet ECG database contains 6 12-lead ECG datasets. However, the data
derive from 4 separate sources.
In total, the ECG data sources are:
\begin{itemize}
	\item \textbf{CPSC, CPSC Extra}. Since both datasets were published during the China Physiological Signal Challenge 2018 (CPSC2018) \cite{liu_open_2018} and originate from the same source, we handle them as a single domain.
	\item \textbf{G12EC}. The second source is the Georgia 12-lead ECG Challenge (G12EC) Database, Emory University, Atlanta, Georgia, USA \cite{alday_classification_2020}.
	\item \textbf{INCART}. The third source is the public dataset from the St. Petersburg Institute of Cardiological Technics (INCART) 12-lead Arrhythmia Database, St. Petersburg, Russia \cite{tihonenko_st-petersburg_2007}.
	\item \textbf{PTB, PTB-XL}. Finally, the PTB datasets were published by Physikalisch-Technische Bundesanstalt (PTB) Database, Brunswick, Germany \cite{wagner_ptb-xl_2020}.
\end{itemize}

As the signals were taken from separate sources, they differ significantly. As a first step, we had to convert them into a common form. Since most of the signals were sampled at 500 Hz and were about 10 seconds long, we resample each signal to 500 Hz and either zero-pad, truncate or split each one into 10-second non-overlapping windows accordingly, to homogenize the input data. Moreover, each ECG recording can be labeled with more than one class. Therefore, this problem falls into the category of multi-label classification. Nonetheless, as the number of present ECG diagnoses is over 100 across all datasets, we followed PhysioNet's guidelines and decided to only classify 24 classes. If none of the signal's labels are present in the 24 classes, we exclude it from the dataset. After this process, 37,749 signal recordings remain in total and each of them is normalized in the $[-1, 1]$ range.

For the EEG setup we adopt 3 SEED datasets, namely SEED, SEED-FRA and
SEED-GER. The above datasets are used for EEG-based emotion recognition
research and contain data from 3 different populations from China, France and
Germany. In all 3 experiments, while presenting the subjects with a series of
film clips chosen to elicit emotions between ``Negative'', ``Neutral'' and
``Positive'', a 62-channel EEG signal, sampled at 1 kHz was captured. After
preprocessing the raw EEG signals, the authors provide the EEG differential
entropy (DE) features which have proven to be more discriminative than the
raw signals \cite{song2018eeg}. The DE features have a dimension of $[N, 5]$, where $N$ is the length of the EEG signal and each of the 5 channels corresponds to a different brain wave frequency, between Alpha, Beta, Gamma, Delta and Theta waves. Hence, we select to use the DE features as input to our models. As the length of each EEG signal varies depending on the length of the provided clip, we once again split the corresponding DE features into 10-second non-overlapping windows. Each window inherits the label from the original signal and is then fed into the model. After splitting the features, we get a total of 852 windows with a dimension of 170 $\times$ 5.

\begin{table*}[h]\centering 
	\begin{center}
		\caption{Per-class results for Intra-distribution and out-of-distribution ECG classification, with a ResNet-$18$ as the backbone network. We compare widely accepted DG algorithms with the baseline (i.e., no DG) model and evaluate each model's performance on intra and OOD data. The top results for each setting are highlighted in $\textbf{bold}$ and are \underline{underlined}, respectively. The total evaluation is based only the per class F1-scores. Additionally, we also report the Macro-F1 scores of each model, when averaged over the classes with at least one true positive example and over the total number of classes in the dataset.}
		\label{tab:ecg_results_res18}
		\begin{adjustbox}{width=1\textwidth}
			\begin{tabular}{|c||cccccc||cccccc|}
				\hline
				Backbone: ResNet-18  & \multicolumn{6}{c||}{Intra-Distribution} & \multicolumn{6}{c|}{OOD} \\
				\hline
				% &  & Intra-Domain &  &  & Intra-Domain &  &  & OOD &  &  & OOD &\\
				% \hline
				%&  & Baseline &  &  & Our Model &  &  & Baseline &  &  & Our Model &\\
				%\hline
				Diagnosis & ERM & IRM & MMD & RSC & CORAL & BioDG & ERM & IRM & MMD & RSC & CORAL & BioDG \\
				%\hline
				\hline
				%~ & ~ & ~ & ~ & ~ & ~ & ~ & ~ & ~ & ~ & ~ & ~ & ~ \\
				1st degree AV block & 67.05 & 0 & 60.87 & 32.33 & 58.35 & \textbf{67.81} & \underline{75.26} & 0 & 71.78 & 44.76 & 70.17 & 70.64 \\
				Atrial fibrillation & 86.14 & 18.18 & 74.21 & 58.89 & 78.76 & \textbf{88.38} & 63.94 & 12.16 & 44.97 & 35.29 & 46.92 & \underline{64.40} \\
				Atrial flutter & 0 & 0 & 0 & 0 & 0 & \textbf{30.00} & 0 & 0 & 0 & 0 & 0 & \underline{4.35} \\
				Bradycardia & 0 & 0 & 0 & 0 & 0 & \textbf{19.05} & 0 & 0 & 0 & 0 & 0 & 0 \\
				Complete right bundle branch block & 77.69 & 16.16 & 68.90 & 72.92 & 66.32 & \textbf{82.41} & 62.35 & 11.50 & 46.92 & 65.42 & 42.83 & \underline{63.97} \\
				Incomplete right bundle branch block & 0 & 0 & 31.97 & 0 & 32.86 & \textbf{47.98} & 0.48 & 0 & \underline{32.74} & 0 & 29.08 & 28.22 \\
				Left anterior fascicular block & 41.59 & 0 & 57.78 & 0 & 54.55 & \textbf{70.44} & 22.22 & 0 & 22.52 & 0 & 19.79 & \underline{39.77} \\
				Left axis deviation & 54.11 & 31.15 & 59.34 & 31.15 & 55.50 & \textbf{72.88} & 26.51 & 17.79 & 38.38 & 17.79 & 35.11 & \underline{40.87} \\
				Left bundle branch block & 81.82 & 0 & 58.63 & 0 & 63.71 & \textbf{88.54} & 67.50 & 0 & 58.22 & 0 & 64.23 & \underline{75.29} \\
				Low QRS voltages & 0 & 0 & 0 & 0 & 0 & 0 & 0 & 0 & 0 & 0 & 0 & 0 \\
				Non-specific intraventricular conduction disorder & 0 & 0 & 0 & 0 & 0 & 0 & 0 & 0 & 0 & 0 & 0 & \underline{5.33} \\
				Pacing rhythm & 87.94 & 0 & 69.90 & 0 & 72.73 & \textbf{91.78} & 0 & 0 & 0 & 0 & 0 & 0 \\
				Premature ventricular contractions & 0 & 0 & 0 & 0 & 0 & 0 & 0 & 0 & 0 & 0 & 0 & 0 \\
				Prolonged PR interval & 0 & 0 & 0 & 0 & 0 & \textbf{15.38} & 0 & 0 & 0 & 0 & 0 & 0 \\
				Prolonged QT interval & 0 & 0 & 0 & 0 & 0 & 0 & 0 & 0 & 0 & 0 & 0 & 0 \\
				Q wave abnormal & 0 & 0 & 0 & 0 & 0 & \textbf{1.71} & 0 & 0 & 0 & 0 & 0 & 0 \\
				Right axis deviation & 0 & 0 & 0 & 0 & 0 & \textbf{30.95} & 0 & 0 & 0 & 0 & 0 & \underline{2.11} \\
				Sinus arrhythmia & 0 & 0 & 0 & 0 & 0 & 0 & 0 & 0 & 0 & 0 & 0 & 0 \\
				Sinus bradycardia & 52.24 & 0 & 0 & 0 & 0 & \textbf{53.01} & 13.76 & 0 & 0 & 0 & 0 & \underline{18.34} \\
				Sinus rhythm & 93.04 & 80.94 & 80.94 & 82.56 & 80.94 & \textbf{93.59} & 52.29 & 31.56 & 31.56 & 31.56 & 31.56 & \underline{50.97} \\
				Sinus tachycardia & \textbf{81.31} & 0 & 80.15 & 0 & 76.68 & 78.22 & 81.49 & 0 & \underline{89.40} & 0 & 89.29 & 79.39 \\
				Supraventricular premature beats & 0 & 0 & \textbf{10.02} & 0 & 6.26 & 5.84 & 0 & 0 & \underline{11.46} & 0 & 9.79 & 1.43 \\
				T wave abnormal & 23.36 & 15.68 & 31.27 & 30.28 & 34.14 & \textbf{32.49} & 1.39 & 38.40 & \underline{38.43} & 29.04 & 36.66 & 1.46 \\
				T wave inversion & 0 & 0 & 0 & 0 & 0 & 0 & 0 & 0 & 0 & 0 & 0 & 0 \\
				%\hline
				\hline
				$\#$ of Predicted Classes & 11 & 5 & 12 & 6 & 12 & \textbf{18} & 11 & 5 & 11 & 6 & 11 & \underline{15} \\ 
				Macro-F1 (Predicted)  & 30.4 & 8.53 & 21.34 & 27.87 & 25.69 & \textbf{48.63} & 23.40 & 7.43 & 17.25 & 11.88 & 20.09 & \underline{35.88} \\ 
				Macro-F1 (All)  & 24.16 & 6.75 & 16.89 & 22.07 & 20.34 & \textbf{38.50} & 14.63 & 4.64 & 10.78 & 7.43 & 12.56 & \underline{22.42} \\
				\hline
			\end{tabular}
		\end{adjustbox}
	\end{center}
\end{table*}

\begin{table*}[h]\centering 
	\begin{center}
		\caption{Per-class results for Intra-distribution and out-of-distribution ECG classification, with a S-ResNet as the backbone network. We compare widely accepted DG algorithms with the baseline (i.e., no DG) model and evaluate each model's performance on intra and OOD data. The top results for each setting are highlighted in $\textbf{bold}$ and are \underline{underlined}, respectively. The total evaluation is based only the per class F1-scores. Additionally, we also report the Macro-F1 scores of each model, when averaged over the classes with at least one true positive example and over the total number of classes in the dataset.}
		\label{tab:ecg_results_small}
		\begin{adjustbox}{width=1\textwidth}
			\begin{tabular}{|c||cccccc||cccccc|}
				\hline
				Backbone: S-ResNet  & \multicolumn{6}{c||}{Intra-Distribution} & \multicolumn{6}{c|}{OOD} \\
				\hline
				
				Diagnosis & ERM & IRM & MMD & RSC & CORAL & BioDG & ERM & IRM & MMD & RSC & CORAL & BioDG\\
				\hline
				%~ & ~ & ~ & ~ & ~ & ~ & ~ & ~ & ~ & ~ & ~ & ~ & ~ \\
				1st degree AV block & 16.14 & 0 & 7.69 & 15.00 & 10.67 & \textbf{61.88} & 7.30 & 0 & 5.84 & 0.51 & 7.38 & \underline{62.24} \\ 
				Atrial fibrillation & 55.84 & 18.18 & 23.98 & 36.99 & 24.81 & \textbf{87.05} & 32.58 & 12.16 & 13.28 & 12.16 & 13.22 & \underline{62.51} \\ 
				Atrial flutter & 19.05 & 0 & 0 & 13.33 & 0 & \textbf{50.00} & 0 & 0 & 0 & 0 & 0 & \underline{10.70} \\ 
				Bradycardia & 0 & 0 & 0 & \textbf{5.26} & 7.69 & 0 & 0 & 0 & 0 & 0 & 0 & 0 \\ 
				Complete right bundle branch block & 73.38 & 16.16 & 60.41 & 72.58 & 64.19 & \textbf{82.46} & 60.15 & 11.50 & 50.90 & 47.75 & 52.64 & \underline{64.55} \\ 
				Incomplete right bundle branch block & 11.01 & 0 & 13.05 & 4.86 & 22.19 & \textbf{43.03} & 14.56 & 0 & 20.09 & 10.85 & 25.65 & \underline{30.27} \\ 
				Left anterior fascicular block & 54.67 & 0 & 50.83 & 52.8 & 53.37 & \textbf{70.29} & 31.03 & 0 & 21.01 & 17.16 & 20.84 & \underline{33.29} \\ 
				Left axis deviation & 58.10 & 31.15 & 58.34 & 56.26 & 60.56 & \textbf{66.87} & 44.98 & 17.79 & 37.75 & 36.09 & 39.16 & \underline{50.42} \\ 
				Left bundle branch block & 80.49 & 0 & 0 & 78.33 & 45.49 & \textbf{85.48} & 62.44 & 0 & 0 & 0 & 40 & \underline{74.04} \\ 
				Low QRS voltages & 0 & 0 & 0 & 0 & 0 & 0 & 0 & 0 & 0 & 0 & 0 & 0 \\ 
				Non-specific intraventricular conduction disorder & 0 & 0 & 0 & \textbf{4.55} & 0 & 2.34 & 0 & 0 & 0 & 0 & 0 & 0 \\ 
				Pacing rhythm & 58.62 & 0 & 37.84 & 43.90 & 48.15 & \textbf{78.46} & 0 & 0 & 0 & 0 & 0 & 0 \\ 
				Premature ventricular contractions & 0 & 0 & 0 & 0 & 0 & 0 & 0 & 0 & 0 & 0 & 0 & 0 \\ 
				Prolonged PR interval & 0 & 0 & 0 & 0 & 0 & \textbf{15.00} & 0 & 0 & 0 & 0 & 0 & 0 \\ 
				Prolonged QT interval & 0 & 0 & 0 & 0 & 0 & 0 & 0 & 0 & 0 & 0 & 0 & 0 \\ 
				Q wave abnormal & 0 & 0 & 0 & 0 & 0 & \textbf{11.11} & 0 & 0 & 0 & 0 & 0 & 0 \\ 
				Right axis deviation & 5.48 & 0 & 23.27 & 9.52 & \textbf{27.92} & 24.53 & 0 & 0 & 18.58 & \underline{21.54} & 17.35 & 2.84 \\ 
				Sinus arrhythmia & \textbf{2.17} & 0 & 0 & 0 & 0 & 0 & 1.94 & 0 & 0 & 0 & 0 & \underline{10.85} \\ 
				Sinus bradycardia & 4.49 & 0 & 0 & 9.03 & 0 & \textbf{36.70} & 31.56 & 31.56 & 31.56 & 31.56 & 31.56 & \underline{47.33} \\ 
				Sinus rhythm & 84.60 & 80.94 & 80.94 & 82.56 & 80.94 & \textbf{91.44} & 38.89 & 0 & 12.54 & 0.62 & 15.17 & \underline{82.68} \\ 
				Sinus tachycardia & 27.86 & 0 & 10.37 & 20 & 15.54 & \textbf{79.82} & 1.35 & 0 & \underline{9.98} & 0 & 8.07 & 1.92 \\ 
				Supraventricular premature beats & 9.27 & 0 & 10.07 & \textbf{11.28} & 8.17 & 3.68 & 0 & 0 & \underline{10.02} & 0 & 6.26 & 5.84 \\ 
				T wave abnormal & 18.57 & 15.68 & 28.62 & 13.31 & 26.14 & \textbf{33.89} & 8.11 &\underline{38.40} & 37.22 & 0 & 30.35 & 4.53 \\ 
				T wave inversion & 0 & 0 & 0 & 0 & 0 & 0 & 0 & 0 & 0 & 0 & 0 & 0 \\ 
				
				\hline
				$\#$ of Predicted Classes & 16 & 5 & 12 & 17 & 15 & \textbf{18} & 12 & 5 & 12 & 9 & 13 & \underline{15} \\ 
				Macro-F1 (Predicted)  & 41.46 & 9.01 & 38.00 & 17.03 & 37.82 & \textbf{53.91} & 31.15 & 7.43 & 32.43 & 14.92 & 31.70 & \underline{36.44} \\ 
				Macro-F1 (All)  & 31.10 & 6.75 & 28.50 & 12.77 & 28.37 & \textbf{40.44} & 19.47 & 4.64 & 20.27 & 9.33 & 19.81 & \underline{22.77} \\
				\hline
			\end{tabular}
		\end{adjustbox}
	\end{center}
\end{table*}

\subsection{Experimental Setup}\label{sec:experimental_setup}

For the DG problem in 1D biosignal data, we move past the leave-one or
leave-multiple-subjects-out protocol and implement the widely accepted in DG,
leave-one or leave-multiple \textit{domains}-out protocol, as described in
\cite{Li_2017_ICCV}. Specifically, in this setting we think of each distinct
dataset, or different data source, as a domain and split all available domains into \emph{Source} and \emph{Target} domains. We train each model on the source
domain data only. Since a DG model should be able to generalize on data from
both source and target domains, we split the model evaluation in Intra-Distribution and OOD evaluation. To that end, for both biosignal datasets, we split the source data into a standard train-val-test split  with a $70-10-20\%$ ratio respectively and use the test split for the Intra-Distribution evaluation. For the OOD evaluation, we use all of the target domain data.

For the ECGs, we split the available databases as follows. The CPSC, CPSC
Extra, PTB and PTB-XL are considered as source domains and the INCART and G12EC
as target domains, following \cite{9898255}. As the number of represented classes differ between datasets, we choose the above target domains to balance the classes in both splits\footnote{The total number of classes in each data source can be found here:
	\url{https://github.com/physionetchallenges/evaluation-2020/blob/master/dx\_mapping\_scored.csv}.}. 
Since the ECG labels are highly unbalanced, we omit reporting results in terms of accuracy (which can be misleading) and instead compare the per-class F1-scores of each method. Note that in some cases, models do not predict \emph{any} true positive for some classes (leading to a zero F1 score). For this reason, we also report the number of classes with at least one true positive example, as an additional indicator of model effectiveness (denoted as ``\# of Predicted Classes'' in the results tables). Using this number, we report the Macro-F1 scores of these predicted classes in addition to the Macro-F1 score for all the classes present in the datasets.

\begin{table}[ht]\centering 
	\begin{center}
		\caption{Intra-distribution and out-of-distribution results for the EEG experiments, with an S-ResNet as the backbone network. We compare widely accepted DG algorithms with the baseline (i.e., no DG) model and evaluate each model's performance on intra and OOD data. The top results for each setting
			are highlighted in $\textbf{bold}$ while the second best are \underline{underlined}. CHI, FRA and GER correspond to data from SEED, SEED-FRA and SEED-GER respectively.}
		\label{tab:eeg_results}
		\begin{adjustbox}{width=1\linewidth}
			\begin{tabular}{|c||ccc|}
				%\multicolumn{7}{|c|}{\hspace{.2cm} Intra-Distribution \hspace{3.5cm} OOD} \\
				\hline
				~ & ~ & Intra & ~ \\
				\hline
				Method & FRA-GER & CHI-GER & CHI-FRA \\
				\hline
				ERM   & 70.94 $\pm$ 2.51 & 73.63 $\pm$ 2.08 & 76.10 $\pm$ 2.27 \\
				IRM   & 70.75 $\pm$ 2.81 & 71.42 $\pm$ 2.98 & 73.90 $\pm$ 3.15 \\
				CORAL & \underline{71.42} $\pm$ 1.67 & 73.72 $\pm$ 1.91 & 74.96 $\pm$ 2.59  \\
				MMD   & 71.23 $\pm$ 2.23 & \underline{74.34} $\pm$ 1.87 & 76.26 $\pm$ 1.95 \\
				RSC   & 70.15 $\pm$ 2.43 & 72.01 $\pm$ 2.37 & \textbf{76.15} $\pm$ 2.08  \\
				\hline
				%\\[-1.ex]
				\textbf{BioDG} & \textbf{77.59} $\pm$ 2.23 & \textbf{76.11} $\pm$ 1.25 & \underline{75.61} $\pm$ 1.15 \\
				\hline
				~ & ~ & OOD & ~ \\ 
				\hline
				Method & CHI & FRA & GER \\
				\hline
				ERM   & \underline{54.60} $\pm$ 3.44 & \underline{45.83}  $\pm$ 3.53  & 63.75 $\pm$ 4.01\\
				IRM   & 53.70 $\pm$ 1.42 & 43.82 $\pm$ 0.75 & 66.87 $\pm$ 3.49 \\
				CORAL & 53.18 $\pm$ 3.29 & 45.00 $\pm$ 2.51 & \underline{67.04} $\pm$ 3.27\\
				MMD   & 54.17 $\pm$ 2.68 & 45.80 $\pm$ 3.46 & 66.04 $\pm$ 3.39 \\
				RSC   & 53.86 $\pm$ 2.55 & 44.37 $\pm$ 2.99 & 58.79 $\pm$ 3.93 \\
				\hline
				\textbf{BioDG} & \textbf{57.09} $\pm$ 0.44 & \textbf{51.74} $\pm$ 0.35 & \textbf{68.04} $\pm$ 0.24 \\
				\hline
				 
			\end{tabular}
		\end{adjustbox}
	\end{center}
\end{table}

\begin{figure}
	\centering
	\includegraphics[width=\columnwidth, ]{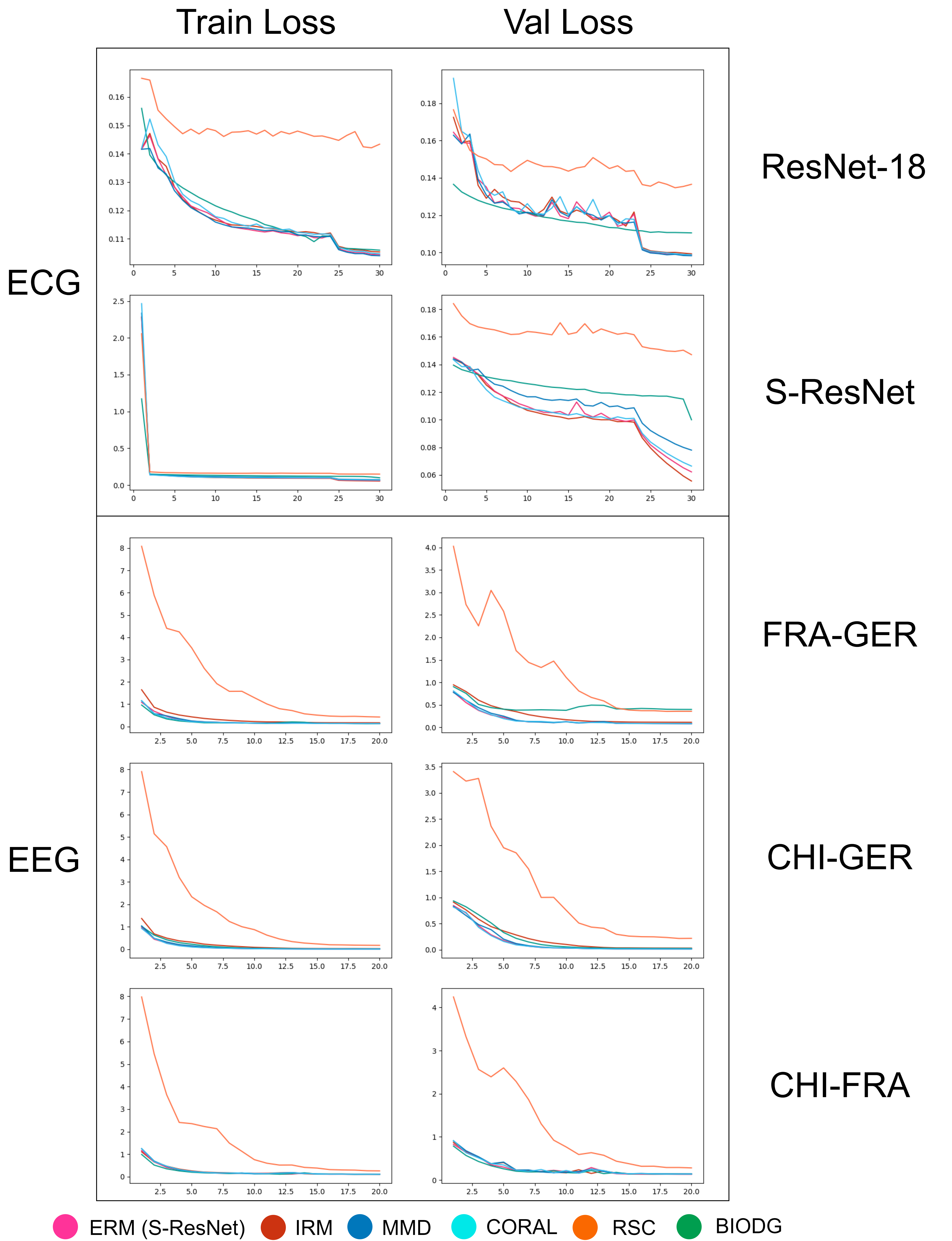}
	\caption{Plots of the training and validation losses of all methods for the BioDG experimental setup. Both training and validation data originate from the \textit{Source} Domains. On the top plots, we illustrate the convergence analysis for the ECG datasets for all models utilizing both the ResNet-18 and S-ResNet backbone networks. The bottom plots depict the training and validation loss progression for the EEG classification task. As in Table \ref{tab:eeg_results}, the labels on the right side of the EEG analysis indicate the (source) domains on which each model was trained on.}
	\label{fig:convergence}
\end{figure}

\begin{figure}[h]
	\centering
	\includegraphics[width=\columnwidth, ]{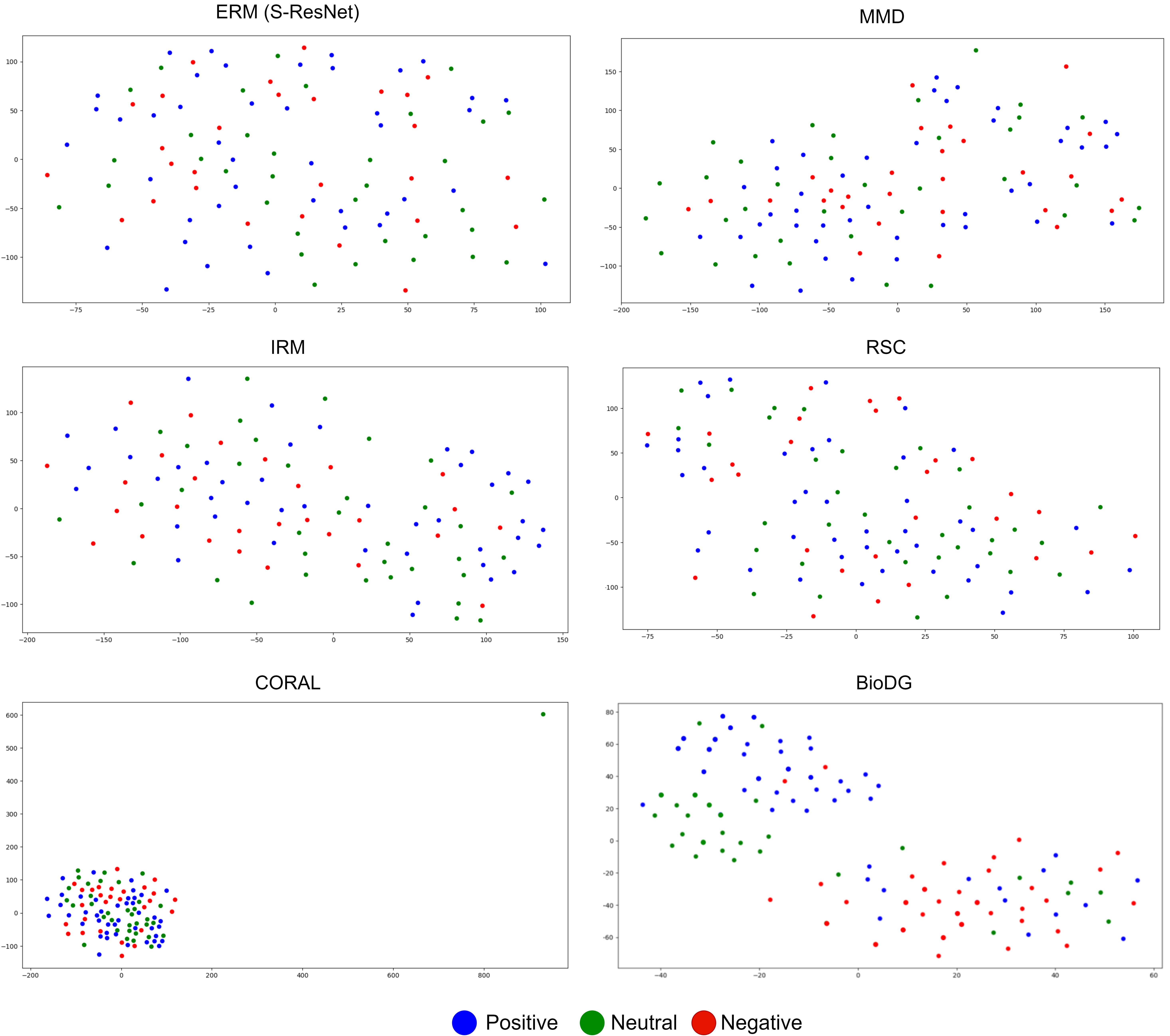}
	\caption{Visualizing the final layer feature vectors of each model in the EEG classification setting using t-Distributed Stochastic Neighbor Embedding (t-SNE). The blue, green and red circles indicate the features which correspond to the Positive, Neutral and Negative emotion classes using EEG signals from the \textit{unseen} test CHI target Domain.}
	\label{fig:tsne}
\end{figure}

For the EEGs, we design three iterations of the experiment based on the
leave-one-domain-out protocol.  In each iteration, we select one of the three
datasets as the target domain and use the remaining two as source data
domains. For example, in the first iteration we select the data from the SEED
dataset as the target domain, originating from China - CHI. Therefore the
intra-distribution evaluation will consist of data from the remaining two
datasets, SEED-FRA and SEED-GER and the OOD evaluation split will contain data
only from CHI. Since this is a multiclass classification problem, we
select to report and evaluate all methods in terms of their total accuracy. 
In our experiments, we repeat each iteration 10 times and present the average 
accuracies, along with the respective standard deviations. The experimental setup for both biosignals is illustrated in Figure \ref{fig:benchmark}.

\section{Results}
\label{sec:results}
For both datasets we implement the baseline models in addition to the
BioDG baseline method, described in Section \ref{sec:ourMethod}. The results for the ResNet-18 and S-ResNet backbones in the ECG experiments are shown in Tables \ref{tab:ecg_results_res18} and \ref{tab:ecg_results_small} respectively, as the results of the EEG experiments 
are presented in Table \ref{tab:eeg_results}. For our experiments, we also provide convergence analysis plots (Fig. \ref{fig:convergence}) of the training and validation losses for all models in both settings, along with t-SNE feature vector visualizations for the EEG biosignal classification task\footnote{We provide t-SNE visualizations only for the the EEG data as the ECG signals are multi-labeled and only a very small subset of them hold a single label and can therefore be attributed to a single feature cluster.}. The experimental results are discussed in the following section.

\subsection{Discussion}\label{discussion}

To ensure the validity of our framework we choose to plot the training and validation loss progression for all methods in each experiment. As illustrated in Figure \ref{fig:convergence}, all algorithms seem to successfully converge on the training and validation data drawn from the source domains. With the exception of RSC in the ECG setting, all algorithms appear to reach about the same validation loss at the end of their training. 
However, when we dive deeper into the experimental results the distribution shift between the source and target domains clearly affects the performance of each model. In the case of the ECG dataset, the overall drop in the Macro F1-score for both evaluation settings is quite obvious. When comparing each model between their intra-distribution and OOD results, it is evident that they are not able to sufficiently generalize to unseen data. However, there is a difference in the performance of each method. In both cases of backbone models the BioDG baseline is consistently one step ahead, as it is able to achieve top performances in both evaluation setups. Furthermore, our method is also able to successfully recognize an additional number of diagnoses in both cases. Another interesting result, is the fact that despite the difference in model parameters (details in Table \ref{complexity}) the performance of the S-ResNet BioDG model is comparable to that of its ResNet-18 counterpart. What's more, in both cases of backbone models the rest of DG algorithms seem to overfit on the sinus rhythm class (normal heartbeat) in contrast to the BioDG baseline. 

The DG problem can also be observed in datasets containing data
from different populations, as in the case of the EEG experiments. Even though
the performance of all models clearly drops when evaluated on unseen data, the proposed BioDG algorithm continues to outperform the rest of the baseline methods. With the exception of the intra-distribution evaluation of CHI-FRA, our network consistently surpasses all other models, in all settings. To provide further evidence of the feature extraction capabilities of our proposed method we visualize the feature vectors of each model's final layer in the EEG setting using t-Distributed Stochastic Neighbor Embedding or t-SNE \cite{maaten_visualizing_2008}. Admittedly, the visualized feature maps are quite erratic and a somewhat clear separation of clusters is not present for any method besides perhaps our own. Even though not all vectors of the same label are grouped together, our algorithm seems to make a more explicit distinction than the baseline models where three clusters appear for the Positive, Neutral and Negative classes.

\begin{table}[h]\centering 
	\begin{center}
		\caption{Complexity metrics of the baseline methods against our proposed BioDG, in both the ECG and EEG settings. We choose to report on the time required for training, memory usage, computational complexity (MACs) and finally, the total number of parameters for each implemented method. Due to the similar complexity metrics of the adapted fully supervised and baseline DG algorithms, we choose to not report on each of individually but as a group and denote them as 'Baselines'.}
		\label{tab:model_complexity}
		\begin{adjustbox}{width=1\linewidth}
			\begin{tabular}{|c|c|c|c|c|}
				\hline
				Methods & Train Time (min.) & Mem Usage (GB) & MACs (G) & Parameters (M)\\
				\hline
				{ECG S-ResNet}  & ~ & ~ & ~ & ~   \\\cline{0-0}
				%\hline
				Baselines   & 12 & 7.33  & 252.253 & 1.755  \\
				\textbf{BioDG}  & 14 & 9.12  & 265.729 & 2.287  \\
				\hline
				
				{ECG ResNet-18}  & ~ & ~ & ~ & ~ \\\cline{0-0}
				%\hline
				Baselines & 21   & 10.32 & 452.405  & 3.894\\
				\textbf{BioDG}  & 23 & 13.81 & 483.262 & 5.178\\
				\hline
				
				{EEG S-ResNet}  & ~ & ~ & ~ & ~ \\\cline{0-0}
				%\hline
				Baselines & 0.5 & 0.20 & 44.301  & 2.019 \\
				\textbf{BioDG}   & 0.5 & 0.25 & 51.124  & 4.431 \\
				\hline
			\end{tabular}
		\end{adjustbox}
	\end{center}
\end{table}

\subsection{Computational complexity}\label{complexity}
Table \ref{tab:model_complexity} provides metrics regarding the computational complexity of the baseline network against our proposed BioDG architecture. As the complexity of the baseline models was comparable, we do not report on each of them separately but instead denote them as `Baselines'. We choose to report on the time needed for training, memory usage, number of parameters and finally, on the total number of multiply-accumulate (MAC) operations of each network. To provide some context, one MAC corresponds to one multiply and addition operation, which can be regarded as equal to 2 FLOPS. 

As expected, the extraction and concatenation of multi-level representations adds a burden to the backbone network. However, since the data are lightweight 1D signals by nature, the additional resources needed is not that high. Even though the number of parameters is greater, the computational complexity (MACs) is increased by around 0.9\% in all cases, which is a relatively small amount. Furthermore, the training time is also comparable, as the only increase is in the ECG experiments by 2 minutes. Finally, the most significant drawback of our method comes from the memory required. With the exception of the EEG S-ResNet model, the ECG S-ResNet and ResNet18 require 1.79 and 3.49 extra GB of memory.

\section{Conclusions}\label{conclusions}

When looking at the above results, it is quite noticeable that as far as DG in
healthcare is concerned, the field has a long way to go. The overarching goal
of this work is to propose an evaluation benchmark for domain generalization in
1D biosignal classification, highlight the importance of additional effort towards the important task of biosignal classification and ultimately prompt further research in the field. By defining structured and reproducible DG experiments, we are able to demonstrate the effect of distribution shifts present in distinct ECG and EEG datasets, when evaluating widely accepted DG algorithms from computer vision. Furthermore, to work towards DG in biosignal classification we attempt to consolidate and leverage representations from across a deep convolutional neural network. We claim that the combination of a CNN's intermediate features can lead to the representation of a biosignal's invariant attributes. The experimental results support our argument, as in most cases our model is able to achieve top performances in both ECG and EEG classification.

As a first step, in the future we aim to extend the evaluation benchmark and include DG setups for multi-dimensional signals and additional downstream tasks, such as medical images and segmentation. In regard to our proposed method, we intend to impose regularization terms and explicitly push the model towards extracting invariant representations of the input signals. Additionally, in conjunction with saliency maps, we also plan on adding attention mechanisms in the concentration pipeline, as an attempt to provide a degree of intuition into the model's inference process.

\bibliographystyle{IEEEtran}
%\bibliography{Library}
\bibliography{references}

%\begin{IEEEbiographynophoto}{Aristotelis Ballas}
%Biography text here without a photo.
%\end{IEEEbiographynophoto}

\begin{IEEEbiography}[{\includegraphics[width=1in,height=1.25in,clip,keepaspectratio]{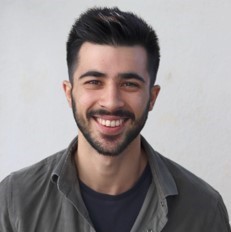}}]{Aristotelis Ballas} is currently working toward the Ph.D. degree in computer science with the Department of Informatics and Telematics, Harokopio University of Athens, Greece. He received his Diploma in 
	Electrical and Computer Engineering from the National Technical University of Athens.
	His research interests include machine learning and representation learning, with an emphasis on domain generalization and AI in healthcare. 
\end{IEEEbiography}

\begin{IEEEbiography}[{\includegraphics[width=1in,height=1.25in,clip,keepaspectratio]{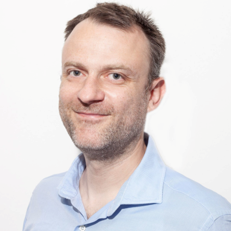}}]{Christos
		Diou} is an Assistant Professor of Artificial Intelligence and Machine
	Learning at the Department of Informatics and Telematics, Harokopio University
	of Athens. He received his Diploma in Electrical and Computer Engineering and
	his PhD in Analysis of Multimedia with Machine Learning from the Aristotle
	University of Thessaloniki. He has co-authored over 80 publications in
	international scientific journals and conferences and is the co-inventor in 1
	patent. His recent research interests include robust machine learning
	algorithms that generalize well, the interpretability of machine learning
	models, as well as the development of machine learning models for the
	estimation of causal effects from observational data. He has over 15 years of
	experience participating and leading European and national research projects,
	focusing on applications of artificial intelligence in healthcare.
\end{IEEEbiography}

\end{document}